\documentclass[leqno,oneeqnum,onefignum,onetabnum,final]{siamart171218}

\usepackage{graphicx}

\usepackage[sc]{mathpazo}
\usepackage{geometry}
\usepackage{graphicx}
\usepackage[sort,compress]{cite}
\usepackage[hyperpageref]{backref}
\usepackage{amsmath,amssymb,mathtools}
\usepackage{graphicx}
\usepackage{siunitx}
\usepackage{paralist}
\usepackage{booktabs}
\usepackage{hyperref}

\newtheorem{issues}{Challenges}
\newtheorem{remark}{Remark}

\graphicspath{{figures/}{tables/}}

\definecolor{red}{RGB}{200,16,46}

\makeatletter
\@ifclassloaded{standalone}{}{
\hypersetup{
linkcolor=red,
urlcolor=red,
citecolor=red}}
\makeatother%

\newcommand{\figref}[1]{Fig.~\ref{#1}}
\newcommand{\tabref}[1]{Tab.~\ref{#1}}
\newcommand{\secref}[1]{\S\ref{#1}}

\newcommand{\mpmri}{mpMRI}

\newcounter{runidnum}

\newcolumntype{R}{>{\columncolor{gray!20}}r}
\newcolumntype{L}{>{\columncolor{gray!20}}l}
\newcolumntype{C}{>{\columncolor{gray!20}}c}

\newcommand{\vect}[1]{\ensuremath{\boldsymbol{#1}}}             
\newcommand{\mat}[1]{\ensuremath{\boldsymbol{#1}}}              

\newcommand{\F}[1]{\ensuremath{\mathcal{#1}}}                   
\newcommand{\D}[1]{\ensuremath{\mathcal{#1}}}                   
\newcommand{\ns}[1]{\ensuremath{\mathbb{#1}}}                   
\newcommand{\fs}[1]{\ensuremath{\mathcal{#1}}}                  
\newcommand{\idiv}{\ensuremath{\nabla\cdot}}                    
\newcommand{\igrad}{\ensuremath{\nabla}}                        

\newcommand{\half}[1]{\frac{#1}{2}}
\renewcommand{\d}[1]{\mathop{}\!\mathrm{d}#1}

\newcommand{\p} {\partial}

\newcommand{\defeq}{\ensuremath{\mathrel{\mathop:}=}}

\newcommand{\T}{\ensuremath{\mathsf{T}}}

\DeclareMathOperator*{\minopt}{minimize}
\DeclareMathOperator*{\maxopt}{maximize}

\newcommand{\bipa}{\begin{inparaenum}[(\itshape i\upshape)]}
\newcommand{\eipa}{\end{inparaenum}}
\newcommand{\bipasub}{\begin{inparaenum}[(\itshape a\upshape)]}
\newcommand{\eipasub}{\end{inparaenum}}

\newcommand{\acr}[1]{{\bf #1}}

\newcommand{\iquote}[1]{``\emph{#1}''}

\newcommand{\mmargin}[1]{{\marginpar{\em\tiny #1}}}\renewcommand{\mmargin}[1]{}

\title{Integrated Biophysical Modeling and Image Analysis: Application to Neuro-Oncology}

\author{%
Andreas Mang\footnotemark[2]\ \textsuperscript{,}\footnotemark[3]
\and Spyridon Bakas\footnotemark[3]\ \textsuperscript{,}\footnotemark[4]\ \textsuperscript{,}\footnotemark[5]\ \textsuperscript{,}\footnotemark[6]
\and Shashank Subramanian\footnotemark[7]
\and Christos Davatzikos\footnotemark[4]\ \textsuperscript{,}\footnotemark[5]
\and George Biros\footnotemark[7]
}

\begin{document}
\maketitle

\renewcommand{\thefootnote}{\fnsymbol{footnote}}

\footnotetext[2]{Department of Mathematics, University of Houston, TX 77204, USA, {\tt andreas@math.uh.edu}}
\footnotetext[3]{Authors contributed equally.}
\footnotetext[4]{Center for Biomedical Image Computing and Analytics (CBICA), University of Pennsylvania, Philadelphia, PA 19104, USA, {\tt sbakas@upenn.edu}, {\tt christos.davatzikos@pennmedicine.upenn.edu}}
\footnotetext[5]{Department of Radiology, Perelman School of Medicine, University of Pennsylvania, Philadelphia, PA 19104, USA}
\footnotetext[6]{Department of Pathology and Laboratory Medicine, Perelman School of Medicine, University of Pennsylvania, Philadelphia, PA 19104, USA}
\footnotetext[7]{Oden Institute of Computational Engineering and Sciences, The University of Texas at Austin, TX 78712, USA, {\tt shashank@oden.utexas.edu}, {\tt gbiros@acm.org}}

\begin{abstract}
Central nervous system (CNS) tumors come with vastly heterogeneous histologic, molecular and radiographic landscape, rendering their precise characterization challenging. The rapidly growing fields of biophysical modeling and radiomics have shown promise in better characterizing the molecular, spatial, and temporal heterogeneity of tumors. Integrative analysis of CNS tumors, including clinically-acquired multi-parametric magnetic resonance imaging (mpMRI) and the inverse problem of calibrating biophysical models to mpMRI data, assists in identifying macroscopic quantifiable tumor patterns of invasion and proliferation, potentially leading to improved \bipa\item detection/segmentation of tumor sub-regions, and \item computer-aided diagnostic/prognostic/predictive modeling\eipa. This paper presents a summary of \bipa\item biophysical growth modeling and simulation, \item inverse problems for model calibration, \item their integration with imaging workflows, and \item their application on clinically-relevant studies\eipa. We anticipate that such quantitative integrative analysis may even be beneficial in a future revision of the World Health Organization (WHO) classification for CNS tumors, ultimately improving patient survival prospects.
\end{abstract}

\section{INTRODUCTION}\label{s:intro}

\begin{figure}
\centering
\includegraphics[width=0.98\textwidth]{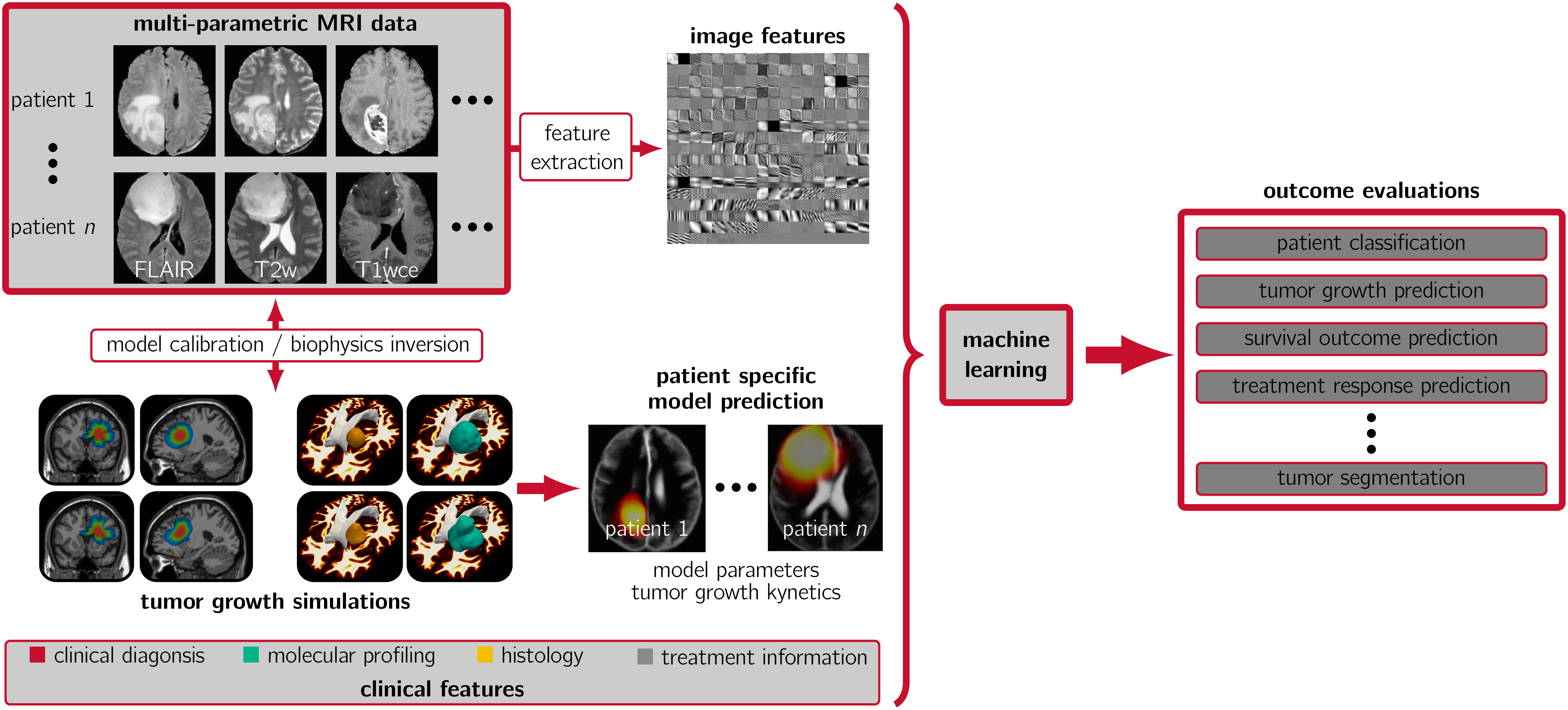}
\caption{Radiomics in neuro-oncology. We seek to extract quantitative imaging indicators that predict clinical outcome. The main inputs to our framework is \mpmri{} data (see top left) and (possibly) clinical features such as molecular profiling and/or histopathological data (bottom left). One possibility to identify clinical markers in imaging data is to apply feature extraction methods from image analysis (top center). These methods do, in general, not incorporate any prior knowledge about the underlying pathology. Computer simulations of biophysical models can establish such a powerful tool to integrate such information. To be clinically useful, biophysical models must be calibrated using the \mpmri{} information (medical images in our case; bottom left). Once calibrated, these models can be used to generate patient specific simulations (bottom center). In a final step, these quantitative parameters are integrated with machine learning algorithms to generate tools that can assist clinical decision-making (right block). (The images included in this figure have been modified from~\cite{Mang:2018a,Gooya:2012a}. Reprinted by permission from Springer Nature, Optimization and Engineering, Copyright \textcopyright2018 Springer and IEEE, IEEE Transactions on Medical Imaging, Copyright \textcopyright2012 IEEE))\label{f:radiomics-overview}}
\end{figure}

Gliomas are the most common primary central nervous system (\acr{CNS}) malignancies. Therapeutic intervention for their most aggressive manifestation---glioblastoma (\acr{GBM})~\cite{Collins:1998, Holland:2000}---remains palliative. Gliomas exhibit highly variable clinical prognosis, and usually contain various heterogeneous sub-regions, with variable histologic and genomic phenotypes. This intrinsic heterogeneity is also characteristic for their radiographic phenotypes---sub-regions appear with different intensity profiles across multi-parametric magnetic resonance imaging (\acr{\mpmri}) scans, reflecting differences in tumor biology and pathophysiology (see \figref{f:radiomics-overview} for an example). Our discussion will be limited to clinical \emph{in vivo} studies in humans; we will not address work in animal models, nor \emph{ex vivo} or \emph{in vitro} studies. Personalized precision medicine aims at developing fine-tuned patient-specific treatment strategies. In the context of neuro-oncology, these include surgery, radiotherapy, and/or chemotherapy planing. Fine-tuning complex clinical treatments necessitates an accurate diagnosis. The fundamental premise that underlies the work of several groups is that biophysical simulations in combination with sophisticated computational methods targeting radiographic features---so called \emph{radiomics}---can augment existing clinical tools, and consequently aid clinical decision-making and patient management.

Current clinical practice is based on the analysis of radiographic imaging data and biopsy, i.e., the \emph{ex vivo} analysis of tissue. Brain tumors have been classified according to the World Health Organization (\acr{WHO}) morphologic-histopathologic classification~\cite{Kleihues:1993a}, from grade I to IV with increasing aggressiveness. In 2016, the WHO has revised its classification scheme into an integrated morphologic-histopathologic and molecular-cytogenetic characterization for CNS tumors~\cite{Louis:2016a} in an attempt to improve tumor stratification potentially leading to an improved patient prognosis. However, even with the addition of molecular-cytogenetic data, CNS tumors---and particularly gliomas---remain challenging to characterize, primarily since their classification is still based on \emph{ex vivo} post-operative tissue analysis (i.e., biopsies).

\begin{issues}[{\bf Shortcomings of Biopsies}]
\noindent Biopsy
$\bullet$ is localized and cannot capture the spatially heterogeneous molecular landscape (\emph{sampling error}),
$\bullet$ is typically not performed longitudinally (i.e., during and after treatment) due to their invasive nature and the potential of neurological deficit (\emph{monitoring limitation}),
$\bullet$ is not feasible for inaccessible, inoperable, and deep-seated tumors (\emph{anatomical constraints}), and
$\bullet$ might be unavailable in many clinical settings due to cost and equipment availability (\emph{economic challenge}).
Despite these shortcomings, tissue analysis provides ground truth and direct cancer molecular information.
\end{issues}

In contrast to tissue analysis, imaging can non-invasively capture \emph{in vivo} the spatial heterogeneity within the whole extent of the tumor (even in deep-seated/inoperable tumors), thereby minimizing potential bias due to only sampling a limited portion of the tumor; moreover it can be performed repeatedly. Since glioma patients routinely undergo multiple \mpmri{} scans---before surgery and during adjuvant treatment---there is ample data available that could help to evaluate the status of the tumor and the surrounding tissue, provide quantitative features for patient assessment, and potentially positively influence personalized treatment and prognosis.

Despite considerable advances in medical imaging sciences, there remains significant challenges. Clinicians face substantial dilemmas during neuroimaging evaluation of patients. For example, for pre-resection patients, a precise quantification of the infiltration of tumor cells into surrounding healthy tissue beyond the visible abnormalities in imaging remains challenging. Differentiation between tumor progression and radiation/treatment effects (a clinical problem termed \emph{pseudo-progression}) can be difficult based on current imaging criteria~\cite{Thust:2018a}; failure to recognize pseudo-progression can lead to premature termination of an effective chemotherapy. On top of that, there exist sensitivities with respect to scanner specific settings and parameters.

In recent years, there has been mounting evidence that quantitative \mpmri{} analysis can characterize CNS tumors comprehensively, and provide critical information about various biological processes within the tumor microenvironment as well as associations with underlying cancer molecular characteristics~\cite{jaffe.radiogenomics, rutman.radiogenomics, mazurowski.radiogenomics, colen.radiogenomics, gevaert.radiogenomics, jain.radiogenomics, itakura.radiogenomics, Elsheikh:2018:gwasGbm, Ellingson:2013:atlasPhenotypes, bakas.ccr, bakas.egfrwt, Rathore:2019:aacr, Rathore:2018:mgmtSno, Rathore:2018:mgmtJco, Binder:2018:SNO, bakas.egfrviii, Bakas:2018:idh, Binder:2018:CancerCell, Beig:2018:radiogenomic, Akbari:2018:egfrviii}. Community efforts have created high-quality datasets that can be used to better understand cancer~\cite{TCIA, tcgaGbmCollection, tcgaLggCollection, Bakas:2018a, AdvancingTCIA, Menze:2015a, tcgaGbmSegmentations, tcgaLggSegmentations, Simpson2019}. Advances in computational inference and machine learning (including deep learning) have dramatically improved our ability to process large datasets. All these advances have facilitated the development of computational methods for high-throughput extraction of quantitative features using sophisticated algorithms. These algorithms are used for image segmentation and can produce quantitative metrics from imaging data, which in turn can be used to produce critical information for patient characterization, especially when fused with other clinical data.

Although purely image-based correlation analysis~\cite{Bakas:2018a} is very successful, there are still many challenges related to robustness (sensitivity to local minima and dataset overfitting) and extrapolation, since most medical datasets are limited compared to the complexity of the underlying processes. Many research studies have sought to extract information using biophysical mode priors in the brain~\cite{Gooya:2012a, Clatz:2005a, Hogea:2007b, Yankeelov:2013a, Konukoglu:2010a, Rockne:2010a, Ivkovic:2012a, Miga:1998a, Lipkova:2019a}, as well as other organs, such as breast, kidney, pancreas, liver, prostate, and lungs~\cite{Chen:2012b, Wong:2015b, Wong:2017a, GarciaCremades:2018a, Filipovic:2014a, Tariq:2015a, Mi:2014a, Prapipkumar:2016a, Atuegwu:2012a, Roque:2018a, Han:2014a, Lorenzo:2016a}. Developments in computational modeling of untreated gliomas, as well as models of polyclonal gliomas following chemotherapy and surgical resection, can help capture important information for diagnostic, planning, and prognostic purposes~\cite{gutman.survival, mazurowski.survival, Bakas:2017:snoAdvancedFeatsAdvancedModalities, bonecamp.survival, akbari.recurrence, nicolasjilwan.survival, velazquez.survival, akbari.recurrence2, batmanghelich.modelling, Rathore:2018:recurrence, macyszyn.survival, rathore.survival, Rathore:2018:beyondIdh, Rathore:2018:SciRep}. The key benefit of these approaches is that they rigorously follow mathematical and physical principles, and are also quantitative and reproducible. They can, in combination with machine learning approaches, help consolidate complex imaging data (see \figref{f:radiomics-overview} for an illustration). They can unveal hidden spatio-temporal variables (i.e., clinical markers that are not directly observable from clinical data). Consequently, the integration of computational models with imaging offers great promise of providing a more complete understanding of clinically relevant entities, thereby improving precision diagnostics and therapeutics. These advances would in turn further improve the clinical outcome and may, ultimately, become an integral part of a new form of WHO classification of CNS tumors. However, developing clinically reliable tumor growth models and their integration with imaging data, which at its core is an inverse problem~\cite{Tarantola:2005a}, remains a significant challenge for various reasons.

\begin{issues}[{\bf Challenges for Integration of Mathematical Models with Imaging}]
\begin{itemize}
\item Tumor dynamics remain mostly unknown; tumor growth is a complex multiscale process that is not entirely understood, and is challenging to capture mathematically. They vary significantly across patients, and across space and time due to differences in the local microenvironment and molecular alterations.
\item It is not possible to conduct controlled experiments that allow for model refinement in humans. Animal models and \emph{in vitro} cultures can help probe different mechanisms but the genome, time scale, and overall environment are quite different in humans. As a result, assessment and validation remains challenging. This issue is further complicated due to therapeutic intervention and resection which are extremely hard to integrate or account for in a simulation-based framework.
\item Mathematical models are typically parameterized by many \emph{unknown} parameters. Calibrating such models requires patient-specific clinical data that is, in general, not available. For example, for GBM patients, most information regarding a tumor's state must be inferred from a single set of \mpmri{} scane (treatment is typically administered immediately after diagnosis).
\item Even if the data were available, the ability to estimate unknown parameters is limited due to fundamental mathematical issues (e.g., non-convexity and ill-posedness of the inverse problem; modeling the observation operator; selecting an appropriate regularization; differentiation and implementation of adjoint equations; noise and uncertainties in data and model; modeling errors).
\item The inverse problem poses computational challenges. If complex models are implemented naively, run times for calibrating them are prohibitive for clinical use. Indeed, even if the forward problem is linear, the inverse problem can be highly nonlinear. As a result, a single calibration can require hundreds of forward problem evaluations. If we consider uncertainty, the costs become even higher.
\end{itemize}
\end{issues}

The significance of the integration of computational models of tumor growth with imaging is three-fold: automatic segmentation of patient images using normal subject images to create spatial (shape) priors; mapping of functional information from atlases to patients (critical in neurosurgery); and parameter calibration of biophysical models. Prior work has shown that biophysical models offer complementary information that relates to tumor aggressiveness and clinical outcome. We summarize some  relevant work and their clinical applications in \tabref{t:clinical-problems}.

\begin{table}
\caption{Clinical problems addressed through integration of mathematical modeling with medical imaging data.\label{t:clinical-problems}}
\centering
\begin{scriptsize}
\begin{tabular}{p{6cm} p{5.5cm}}\toprule
Clinical Problem & References \\\midrule
{Tumor grading and profiling} & \cite{macyszyn.survival, Rathore:2018:beyondIdh, Rathore:2018:SciRep} \\
{Molecular characterization} & \cite{Elsheikh:2018:gwasGbm, Rathore:2019:aacr, Rathore:2018:mgmtSno, Rathore:2018:mgmtJco, Binder:2018:SNO, Bakas:2018:idh, Binder:2018:CancerCell, Akbari:2018:egfrviii}\\
{Growth prediction} & \cite{Chen:2012b, Wong:2015b, Wong:2017a, Prapipkumar:2016a, Benzekry:2014a, Collis:2017a, Hormuth:2015a, Hormuth:2017a, Oden:2010a, Rahman:2017a, Wong:2015a} \\
{Infiltration margins (surgical planning)} & \cite{Rekik:2013a, Baldock:2014:survivalBenefit, Mosayebi:2012a} \\
{Planning of radiotherapy} & \cite{Rockne:2010a, Lipkova:2019a, Corwin:2013a, Le:2017a, Unkelbach:2014a} \\
{Prognosis \& survival prediction} & \cite{GarciaCremades:2018a, Bakas:2017:snoAdvancedFeatsAdvancedModalities, macyszyn.survival, rathore.survival, Swanson:2008a, Wasserman:1996a} \\
{Tumor recurrence prediction} & \cite{akbari.recurrence, akbari.recurrence2, Rathore:2018:recurrence} \\
{Prediction \& modeling of treatment response} & \cite{Yankeelov:2013a, Rockne:2010a, Tariq:2015a, Mi:2014a, Baldock:2014:survivalBenefit, Jackson:2015a, Lima:2017a, Powathil:2007a, Weis:2017a}\\
{Improvement of imaging workflows} & \cite{Gooya:2012a, Bakas:2018a, AdvancingTCIA, Scheufele:2018a, Zacharaki:2009a, Stefanescu:2004a, Angelini:2007a, Bauer:2013a, Bakas:2016a, Mi:2015a, Hogea:2008a, Mang:2012c} \\
\bottomrule
\end{tabular}
\end{scriptsize}
\end{table}

\paragraph*{Contributions} In this manuscript, we \bipa\item review state-of-the-art approaches in tissue level brain tumor modeling, \item present mathematical strategies for model calibration, \item discuss the integration of biophysics simulations with medical imaging data to aid imaging workflows and, ultimately, generate predictive capabilities, and \item showcase different clinical studies that benefit from such an integration\eipa.

\paragraph*{Outline} In the present work, we focus on forward simulation of tumor growth on a macroscopic scale (tissue level), and inverse problem formulations that connect sophisticated forward models with imaging methods (see \figref{f:radiomics-overview}). We review and present formulations and methodology for the simulation of tumor growth in \secref{s:tumor-modeling}. We describe approaches for model calibration and biophysics inversion in \secref{s:inversion}. We discuss the integration of biophysics simulation and computational methods for radiomics in \secref{s:integration-with-mri}. We provide results from clinically relevant studies in \secref{s:clinically-relevant-studies}.

\section{TUMOR GROWTH MODELING AND SIMULATION}
\label{s:tumor-modeling}

There is a long tradition in the design of mathematical models of tumor progression~\cite{Bellomo:2008a,Roose:2007a,Wang:2008a,Harpold:2007a}. Recent advances in mathematics and computational engineering have led to a rich pool of computational models with unprecedented complexity. These models present us with significant challenges; they encompass multiscale, strongly heterogeneous, and coupled multiphysics behavior. Models range from simple population growth models~\cite{Benzekry:2014a} to complex multiphysics, multispecies, space-time models~\cite{Rahman:2017a,HawkinsDaarud:2013a,Gu:2012a}, with dynamical systems that describe tumor progression on various scales of observation, including molecular~\cite{Schuetz:2013a}, cellular~\cite{Toma:2012e,Toma:2013a}, tissue~\cite{Swanson:2000a,Swanson:2002a,Swanson:2003a}, or multiscale representations~\cite{Deisboeck:2011a,Lowengrub_2009,Schuetz:2014a}. We will limit ourselves to models that can be integrated with \emph{in vivo} morphological or functional medical imaging such as MRI, CT, or PET (i.e., models that yield outputs on a tissue scale). Cancer progression is typically formulated as a dynamical system (a set of ordinary (\acr{ODE}s) or partial differential equations (\acr{PDE}s)) based on principles of conservation and constitutive laws. For tissue-level models, tumor cells are not tracked individually but modeled as a concentration or volume fraction (assuming constant density) $c(\vect{x},t)$, where $\vect{x} \in \Omega \subset \ns{R}^3$ and $t\in [0,1]$ (where $t$ has been non-dimensionalized to the unit interval). Depending on the model, $c$ can be a scalar (single species) or a vector (multiple species).

\begin{figure}
\centering
\includegraphics[width=0.9\textwidth]{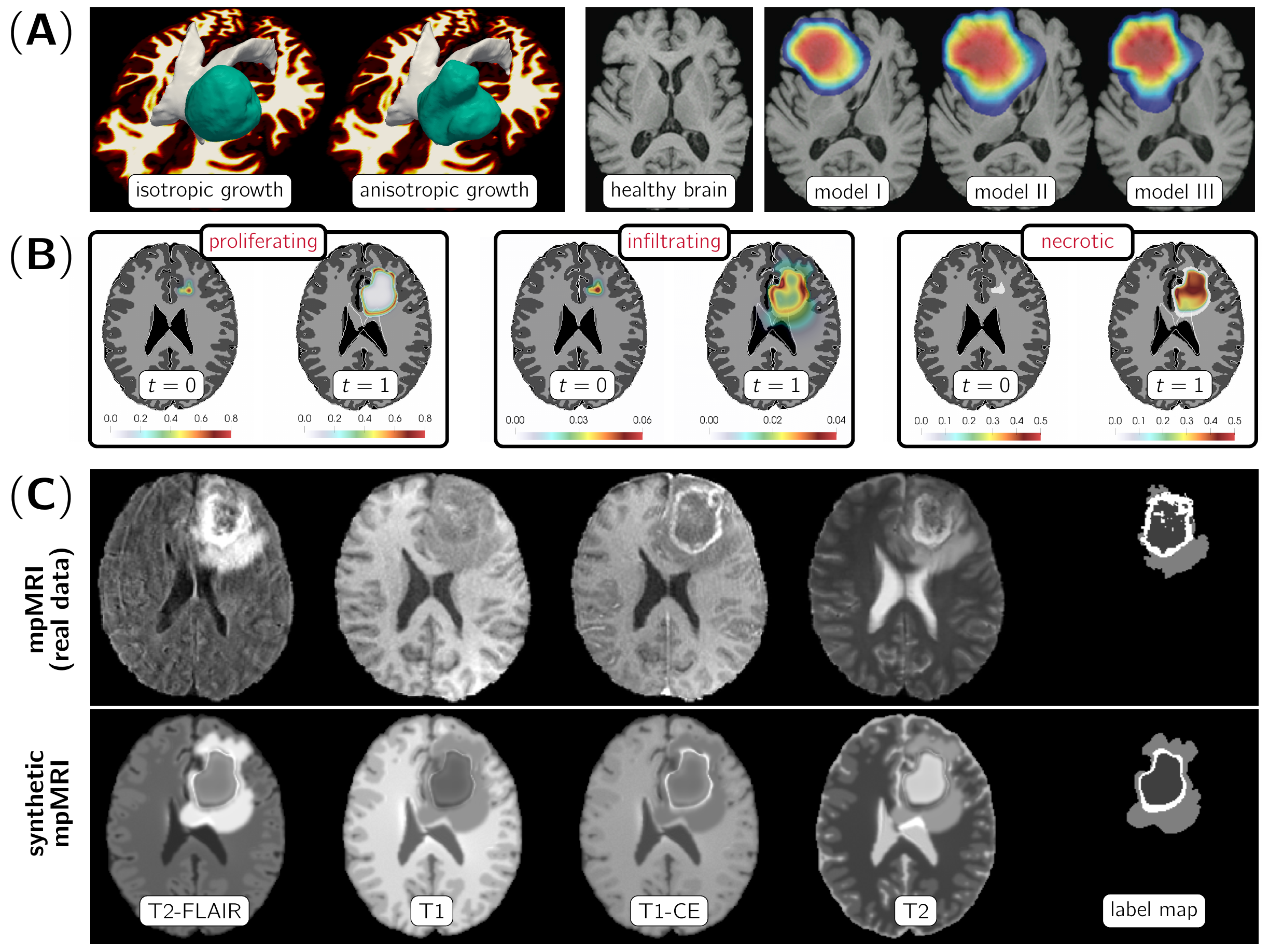}
\caption{Qualitative simulation results for different biophysical models. ({\bf A}) Two single species models, one without (left) and one with (right) mass effect. The three images on the right show results (axial slices through the brain) for different realizations of a mass effect model; we show different degrees of deformation of the healthy tissue due to tumor growth~\cite{Hogea:2008a}. ({\bf B}) Simulation results for a multispecies model of tumor growth with mass effect~\cite{Subramanian:2019a}. We show two time points (initial condition and final time) per tumor species (left to right: proliferating, infiltrating, and necrotic tumor cells). ({\bf C}) This multispecies model allows us to account for imaging abnormalities seen in \mpmri. Top row: patient-specific \mpmri{} data for a GBM; bottom row: synthetically generated \mpmri{} dataset using the model described in~\cite{Subramanian:2019a}. The model parameters were identified by manual trial and error; no inversion was performed. (The simulation results included in this figure have been modified from~\cite{Mang:2018a,Hogea:2008a,Mang:2012b,Subramanian:2019a}. Figure ({\bf A}) left: Reprinted by permission from Springer Nature, Optimization and Engineering, Copyright \textcopyright2018 Springer. Figure ({\bf A}) right: Copyright \textcopyright2008 Society for Industrial and Applied Mathematics. Reprinted with permission. All rights reserved. Figure ({\bf B}) and Figure ({\bf C}): Reprinted by permission from Springer Nature, Journal of Mathematical Biology, Copyright \textcopyright2019 Springer.)\label{f:forward-models}}
\end{figure}

The seminal works~\cite{Swanson:2000a,Swanson:2002a,Swanson:2003a} are based on the assumption that cancerous cells either originate from cell division (proliferation) or cell migration. These principles can be captured by reaction-diffusion (\acr{RD}) equations of the form
\begin{equation}
\label{e:fwd:rd}
\p_t c - \kappa \D{D} c  - f(c) = 0 \text{ in } \Omega \times (0,1],
\qquad c = c_0 \text{ in } \Omega \times \{0\},
\end{equation}

\noindent with zero flux boundary conditions on $\p\Omega$ and initial condition $c_0$ at $t=0$.  $\D{D}$ is a diffusion operator that models the migration of cancerous cells into surrounding healthy tissue, parameterized by the diffusion coefficient $\kappa \geq 0$. The functional $f$ models the proliferation of tumor cells parameterized by $\rho \geq 0$; the most common model is a logistic growth function $f(c) =  \rho c(1 - c)$. Further, $\D{D}c \defeq \idiv \mat{K} \igrad c$ with gradient $\igrad \defeq(\p_1,\ldots,\p_d)$, divergence operator $\idiv \defeq \sum_{i=1}^d \p_i$, and $\mat{K} : \bar{\Omega} \to \ns{R}^{d\times d}$. The tensor field $\mat{K}$ controls the diffusion within different tissue compartments. Initial models~\cite{Swanson:2000a,Swanson:2002a,Swanson:2003a} considered distinct diffusion rates in white matter (\acr{WM}) and gray matter (\acr{GM}). These models were extended to account for a preferential (anisotropic) diffusion within WM by integrating diffusion tensor imaging data~\cite{Mang:2014a,Clatz:2005a,Konukoglu:2010a,Gholami:2016a,Jbabdi:2005a,Mang:2012b,Konukoglu:2010b}. 

While the single species model~\eqref{e:fwd:rd} can capture the overall dynamics of tumor growth, it  does not capture the imaging phenotype of gliomas. For example, GBM typically presents with an enhancing rim surrounding a necrotic core, with significant peritumoral edematous/tumor-infiltrated tissue (see \figref{f:radiomics-overview} for an example). Also, it does not capture the mechanical deformation of the brain parenchyma, the so-called \iquote{mass-effect}~\cite{Clatz:2005a,Chen:2012b,Wong:2017a,Hormuth:2017a,Wong:2015a,Weis:2017a,Hogea:2008a,Mohamed:2005a,Hogea:2007a,Subramanian:2019a}. Models that attempt to capture the heterogeneous phenotype use multiple species of tumor cells with different underlying hypotheses that govern their evolution (see \figref{f:forward-models}). One class of models assumes that tumor cells exist in interchangeable states based on the nutritional condition of their environment. A popular hypothesis is \iquote{go-or-grow}, which stipulates that invading tumor cells are minimally proliferative and vice-versa~\cite{hatzikirou2012}. Models that follow this hypothesis represent tumor progression as (a cycle of) two phases: an initially exclusively proliferative phase followed by an invasion of tumor cells into surrounding tissues. This second phase can then possibly transition back to a proliferating phenotype, which encourages recurrence and growth of metastatic tumors. Other models~\cite{Swanson:2011a} consider these phases to be difficult to isolate; they are modeled to occur simultaneously. As an example, consider a multispecies model that accounts for mass-effect and in which $\vect{c}(\vect{x},t) \defeq (c_{\text{P}}(\vect{x},t), c_{\text{I}}(\vect{x},t), c_{\text{N}}(\vect{x},t))$ consists of proliferating ($\text{P}$), invading ($\text{I}$), and necrotic ($\text{N}$) tumor cell phenotypes, respectively. The associated system of PDEs (including mass-effect) is given by~\cite{Subramanian:2019a, saut2014multilayer}
\begin{subequations}
\label{e:fwd:multispecies}
\begin{align}
\p_t \vect{c} + \idiv (\vect{c}\otimes\vect{v}) - \kappa\D{D}\vect{c} - \vect{f}(\vect{c}) - \vect{g}(\vect{c},\vect{n}) & = \vect{0}                    && \Omega \times [0,1] \\
                                                 \p_t \vect{n} - \D{D} \vect{n} - \vect{h}(\vect{c}, \vect{n}, \vect{m}) & = 0                           && \Omega \times [0,1] \\
                                            \idiv (\lambda \igrad \vect{u} + \mu (\igrad \vect{u} + \igrad\vect{u}^\T))  & = \vect{b}(\vect{c})          && \Omega \times [0,1] \\
                                                                                                           \p_t \vect{u} & = \vect{v}                    && \Omega \times [0,1] \\
                                                                        \p_t \vect{m} + \idiv (\vect{m} \otimes\vect{v}) & = \vect{s}(\vect{m},\vect{c}) && \Omega \times [0,1]
\end{align}
\end{subequations}

\noindent with initial conditions $c = c_0$ and $\vect{m} = \vect{m}_0$; $\otimes$ denotes the outer product, $\vect{u}(\vect{x},t)$ is the displacement field, and $\vect{v}(\vect{x},t)$ is the derived velocity field.

First, notice the two-way coupling between mass-effect and tumor growth due to an advective term in the RD equation and a forcing term in the linear elasticity term. Second, the mass effect is modeled using a simple linear elasticity model with a forcing proportional to the gradient of $c$. More complex models that account for large deformations, growth stresses, residual stress, and tissue microstructure exist~\cite{ambrosi2011perspectives,goriely2015mechanics,Rutter:2017:darcytumorgrowth}.  The linear elasticity model is parameterized by the inhomogeneous Lam\'e coefficients $\lambda : \bar{\Omega}\to\ns{R}$ and $\mu : \bar{\Omega}\to\ns{R}$; they depend on the underlying tissue type. The function $\vect{b} \propto \igrad c$ represents a body force acting on the brain parenchyma~\cite{Clatz:2005a,Wasserman:1996a,Hogea:2008a,Subramanian:2019a}. Finally, the last two equations in~\eqref{e:fwd:multispecies} model the evolution of healthy tissue volume fractions $\vect{m}(\vect{x},t) \defeq (m_{\text{W}}(\vect{x},t), m_{\text{G}}(\vect{x},t), m_{\text{F}}(\vect{x},t))$, where W, G, and F designate WM, GM, and cerebrospinal fluid ({\bf CSF}), and initial condition $\vect{m}_0(\vect{x}) \defeq (m_{\text{W},0}(\vect{x}), m_{\text{G},0}(\vect{x}), m_{\text{F},0}(\vect{x}))$. The field $\vect{s}(\vect{m},c)$ models the rate of change of tissue due to sources and sinks. For example, it can account for CSF leakage and/or loss of healthy GM and/or WM cells due to tumor progression. We illustrate qualitative simulation results for this model in~\figref{f:forward-models}. In addition to the spatiotemporal dynamics of tumor cells, we also account for nutrient supply and tumor-induced edema (\acr{ED}). These quantities are represented as a vector-valued function $\vect{n}(\vect{x},t) \defeq (n_{\text{O}}(\vect{x},t), n_{\text{E}}(\vect{x},t))$ consisting of two concentration maps for nutrients/oxygen ($\text{O}$) and ED ($\text{E}$), respectively. Changes in $\vect{n}$ are modeled via a diffusion operator $\D{D}$ and a source and/or sink term $\vect{h}(\vect{c}, \vect{n}, \vect{m})$. The sources and sinks can, for example, account for the supply and consumption of oxygen, and the leakage of ED into the extracellular matrix due to migrating tumor cells. The RD equation for the tumor phenotypes $\vect{c}$ includes a sink and/or source term $\vect{g}(\vect{c}, \vect{n})$. The precise form of $\vect{g}$ depends on the underlying hypotheses of the growth model. The \iquote{go-or-grow} hypothesis, for example, stipulates mutually incompatible proliferating and migrating phenotypes through different reaction and diffusion operators. The transition between phenotypes depends on the local oxygen concentration. The mass effect is modeled using a linear elasticity equation. We illustrate qualitative simulation results for this model in~\figref{f:forward-models}.

Despite their phenomenological character, these types of models (in particular the single species model coupled with mass effect) can successfully capture the overall appearance of tumors in \mpmri~\cite{Mang:2014a,Clatz:2005a,Konukoglu:2010a,Hormuth:2015a,Rekik:2013a,Mang:2012b,Liu:2014a}. They have been used to \bipa\item study tumor growth patterns in individual patients~\cite{Mang:2014a,Clatz:2005a,Rekik:2013a,Swanson:2000a,Mang:2012b}, \item extrapolate the physiological boundary of tumors~\cite{Konukoglu:2010a,Mosayebi:2012a}, or \item study the effects of clinical intervention~\cite{Rockne:2010a,Mi:2014a,Le:2017a,Swanson:2008a,Powathil:2007a,Weis:2017a,Le:2015a}\eipa. Applications for these types of models beyond brain tumor imaging are, e.g., breast~\cite{Weis:2017a} and pancreatic cancer~\cite{Liu:2014a}. Continuum models of the form~\eqref{e:fwd:rd} have also been extended to account for the evolution of cancer progression on the cellular scale (e.g., accounting for healthy, proliferative, quiescent, and necrotic cellular phenotypes), the subcellular scale, and the molecular scale (for example, accounting for signaling pathways)~\cite{Rahman:2017a}.

\section{INVERSE PROBLEMS FOR PARAMETER CALIBRATION}
\label{s:inversion}

Next, we discuss the inverse problem of estimating biophysical model parameters, $\vect{p}$, for a given tumor growth model, $\D{F}$, with the ultimate goal to provide a framework for patient-specific tumor growth simulations and model predictions. A natural approach to estimate $\vect{p}$ is to formulate a PDE-constrained optimization problem.

\begin{remark}
We only discuss the problem of estimating tumor specific parameters. The integration of tumor modeling with \mpmri{} is described in \secref{s:integration-with-mri}. In practice, imaging is used to derive information for calibrating the model. For example, $c_{\text{OBS}}$, which we will define in the next paragraph, is implicitly derived from \mpmri{} data.
\end{remark}

\subsection{Deterministic Formulations}

The input to our problem are a series of observations of tumor cell densities $c_{\text{OBS}}$ (partially observed tumor data) at specified time instances $\{t^j\}_{j=1,\ldots,n_t}$ within a given time interval $[0,T]$ with final time $T > 0$. In the inverse problem, we seek parameters $\vect{p}$ such that the model output $c(\vect{x},t)$ (i.e., the simulated cell density or tumor cell probability maps) matches these observations. In a general format, we can represent this calibration of tumor models as a PDE-constrained optimization problem of the form
\begin{equation}\label{e:pdeopt}
\minopt_{\vect{p}} \; \half{1}\sum_{j=1}^{n_t}\int_0^T\!\!\!\!\int_{\Omega}\!\!\delta(t - t^j)\big(c(\vect{x},t) - c^j_{\text{OBS}}(\vect{x})\big)^2\d \vect{x}\d t  + \fs{R}(\vect{p})
\;\text{subject to}\; \D{F}(\vect{p},c) = 0.
\end{equation}

\noindent This formulation balances the data fidelity with regularity assumptions on the model parameters $\vect{p}$ (the inversion variable of our problem). We consider a squared $L^2$-distance to measure the discrepancy between $c_{\text{OBS}}^j(\vect{x})$ and the model output $c(\vect{x},t)$. The operator $\D{F}$ is the forward tumor model (see \secref{s:tumor-modeling} for examples). The \emph{inversion variable} $\vect{p}$ (as an example, the growth rate $\rho\geq0$) and the \emph{state variable} $c$ (as an example, the density of tumor cells) are the unknowns of our problem. $\delta$ is a Dirac delta function to pick the time points $t^j$ in $[0,T]$ to which the data $c^j_{\text{OBS}}$ is associated. The functional $\F{R}$ in \eqref{e:pdeopt} is introduced to alleviate the ill-posedness of the inverse problem of recovering $\vect{p}$ from $c_{\text{OBS}}$. The basic idea is to stably compute a locally unique solution to a nearby problem by imposing prior knowledge based on an adequate regularization scheme (in our case, represented by $\F{R}$).

\begin{remark}
In~\eqref{e:pdeopt}, we assume that we are given a time-series $c_{\text{OBS}}^j$, $j=1,\ldots,n_t$. In a typical clinical setting, we are usually given a single snapshot for $c_{\text{OBS}}$. Indeed, when a patient presents with symptoms, the tumor is usually large and treatment (surgery, chemotherapy, or radiation) starts immediately. Consequently, it is not practical to assume a time series of data; any treatment that takes place needs to be incorporated into the models, which poses additional difficulties. This is different in animal studies (which we do not consider here). We discuss strategies to resolve this issue of limited data in \secref{s:integration-with-mri}.
\end{remark}

The standard approach for solving~\eqref{e:pdeopt} is to introduce an additional unknown---the Lagrange multiplier $\lambda$---and derive the stationarity conditions for the Lagrangian $\F{L}(c,\vect{p},\lambda) \defeq \F{J}(\vect{p}) + \langle\D{F}(c,\vect{p}),\lambda\rangle_{L^2(\Omega)}$ ($\F{J}$ denotes the objective functional in~\eqref{e:pdeopt}). Derivative-free optimization~\cite{Mang:2014a,Chen:2012b,Wong:2015b,Wong:2017a,Mi:2014a,Hogea:2008a,Mang:2012b,Konukoglu:2010b,Hogea:2007a,Knopoff:2017a}, automatic differentiation, or finite-difference approximation of the gradient~\cite{Hormuth:2015a} are other options. Derivative-free strategies are easy to implement (they only require routines to evaluate $\D{F}$ and $\D{J}$ for trial parameters $\vect{p}$). However, they lead to sub-optimal algorithms with slow convergence, typically resulting in an excessive number of iterations and high computational costs (perhaps run times of days on medium size clusters). This renders these methods impractical, especially for problems parameterized by a large number of unknowns $\vect{p}$. The works in~\cite{Hogea:2008a,Gholami:2016a,Liu:2014a,Knopoff:2017a,Colin:2014a,Feng:2018a,Knopoff:2013a,Quiroga:2015a} use adjoint information, i.e., methods that exploit analytical derivatives. These methods are preferable to derivative-free approaches as they offer better convergence guarantees, are founded on rigorous mathematical principles, can reveal structure (sensitivities) that can be rigorously analyzed, and can be exploited for further integration with imaging (e.g., construction of priors for Bayesian inference).

A variety of different approaches and formulations have been considered. In~\cite{Knopoff:2013a}, the tumor is modeled as a radially symmetric spheroid. More complex tumor models are described in~\cite{Wong:2015a, Hogea:2008a, Liu:2014a}. In~\cite{Hogea:2008a}, adjoint equations in 1D are derived; derivative-free optimization is used for 3D. The work in~\cite{Liu:2014a} extends on~\cite{Hogea:2008a}; they provide results for an adjoint based method for 3D problems. The work in~\cite{Wong:2015a} follows up on~\cite{Hogea:2008a,Liu:2014a}; the key difference is that they use a hyperelastic mass-effect model as opposed to linear elasticity~\cite{Hogea:2008a,Liu:2014a}. Complex multispecies models are considered in~\cite{Colin:2014a, Quiroga:2015a}. An inversion framework to determine the initial distribution of tumor concentration of a non-linear RD PDE using adjoint-based methods is described in~\cite{Scheufele:2018a,Gholami:2016a,Gholami:2017a}.

All works discussed above calibrate only for a subset of unknown parameters $\vect{p}$. For example, in~\cite{Scheufele:2018a} the authors assume a known reaction coefficient $\rho$ and invert for the tumor initial condition $c_0(\vect{x})$ and a scalar diffusion coefficient $\kappa > 0$. Indeed, it is not possible to reliably invert for all parameters---even for simple RD PDEs---due to the ill-posedness of the inverse problem~\cite{Konukoglu:2010b,Subramanian19aL1}. Additional modeling assumptions can alleviate some of these issues. In~\cite{Rekik:2013a}, a localized initial condition is estimated along with the tumor diffusion coefficient for a traveling wave approximation. This localization enables the estimation of more biophysically meaningful diffusion rates. The work in~\cite{Jaroudi2019} attempts to reconstruct sparse tumor initial condition while fixing the other parameters of a 3D non-linear RD model. The work in~\cite{Subramanian19aL1} describes a framework to estimate all unknown parameters of a 3D non-linear RD growth model, i.e., the reaction coefficient, the diffusion coefficient, and the tumor initial distribution. Sparsity constraints on the tumor initial conditions and constraints on its maximum norm are introduced. This work extends~\cite{Scheufele:2018a} with a greedy pursuit algorithm for imposing sparsity constraints.

\subsection{Probabilistic Formulations}

While the deterministic approach described above is adequate for identifying optimal parameters $\vect{p}^\star$ such that model predictions match some observed data $c_{\text{OBS}}$, its practical value remains limited. Indeed, solving~\eqref{e:pdeopt} only provides us with point estimates $\vect{p}^\star$. In practice, we are interested in predicting some future quantity of interest $q(\vect{p})$, say, the probability of tumor recurrence after surgery. Due to uncertainties in the data $c_{\text{OBS}}$, the inversion variable $\vect{p}$, and the mathematical model $\D{F}$, as well as the non-convexity of the inverse problem, we require confidence intervals for $q(\vect{p})$, and not just point estimates $\vect{p}^\star$. This can be achieved through a probabilistic formulation of the inverse problem, the result of which is a probability density function (\acr{PDF}) that characterizes our confidence in $q(\vect{p})$. The appropriate framework for dealing with such problems is Bayesian inference~\cite{Tarantola:2005a}. Bayesian inference yields a systematic framework that rigorously follows mathematical and physical principles and enables us to address key questions underlying predictive computational modeling. We can quantify uncertainties as they propagate through all steps of our system and we can assess model validity and adequacy. While these features are appealing, it must be noted that a serious drawback is the significant increase in computational burden. Approaches that consider a probabilistic framework are described in~\cite{Collis:2017a,Oden:2010a,Le:2017a,Lima:2017a,Le:2015a,HawkinsDaarud:2013b,Menze:2011a,Oden:2013a,Paek:2014a}. The works in~\cite{Le:2015a,Menze:2011a,Meghdadi:2017a} are based on RD-type systems. The works in~\cite{Oden:2010a,Lima:2017a,HawkinsDaarud:2013b,Oden:2013a} present methodology for statistical model calibration and, in addition to that, provide methodology and results for model selection and validation. They, similarly to~\cite{Kahle:2018a}, consider phase-field models. An extension of~\cite{Oden:2010a,HawkinsDaarud:2013b,Oden:2013a} is described in~\cite{Lima:2017a}; here, a set of eight RD-type models and five variants of phase-field models of varying complexity are considered. The works in~\cite{Collis:2017a,Paek:2014a,Patmanidis:2018a} are restricted to simplistic models of spheroid tumor growth. The focus is on developing effective stochastic computational methods. The works mentioned so far focus mostly on algorithmic developments. The work in~\cite{Kahle:2018b} discusses theoretical considerations, instead. It is an extension of~\cite{Kahle:2018a} to the probabilistic setting. They, likewise to~\cite{Oden:2010a,HawkinsDaarud:2013b,Oden:2013a}, consider a phase-field model. Well-posedness results of the posterior measure for general prior measures are provided.

\begin{issues}[{\bf Challenges and Open Questions in Biophysical Inversion}]
\noindent Inverse problems pose not only computational challenges but also additional modeling challenges that have not been addressed thoroughly in existing work. Key questions are: $\bullet$ What is an appropriate mismatch function to quantify the discrepancy between model prediction and imaging data? $\bullet$ What are appropriate regularization models for $\vect{p}$? $\bullet$ How do numerical schemes affect the computed solutions? $\bullet$ How to efficiently deal with the ill-posedness of the inverse problem? $\bullet$ How to deal with sparse/scarce data? Although general frameworks from other disciplines do exist, specializing these techniques to the specific problem is critical for the clinical success of inverse modeling. Developing adequate methods for inverse modeling remains an open problem for biophysical models of tumor growth.
\end{issues}

\section{INTEGRATION WITH MRI}
\label{s:integration-with-mri}

The integration of biophysical tumor growth models with \mpmri{} can be considered as a two-way coupled problem, where imaging provides the data required to drive the calibration of a biophysical model through the solution of an inverse problem, and biophysical models can define priors for image analysis and introduce additional biomarkers. Imaging data for calibration include the geometry of the brain and the implicit characterization of the underlying tissue, via intensity information from \mpmri{} modalities. Biophysical models provide probabilistic information of tumor infiltration in specific tissues to, e.g., enable or guide common imaging workflows, such as image segmentation~\cite{Gooya:2012a, Binder:2018:CancerCell, Scheufele:2018a, Bakas:2016a, Prastawa:2009a} or image registration~\cite{Zacharaki:2009a, Hogea:2008a, Mohamed:2006a, Kwon:2014a}. Biophysical model parameters can also play the role of biomarkers, for example the reaction and diffusion coefficients, and parameters of the initial condition such as focality and location.

\subsection{Imaging Workflows}

In \emph{image segmentation}, we seek a labeling of the medical imaging data into different sub-regions, each of which corresponds to tissue with distinct pathophysiological properties. In our application, we are interested in differentiating healthy and diseased brain tissue, and possibly subdivide further the healthy and non-healthy regions. In neuro-oncology for high-grade gliomas, one typically differentiates the anatomical brain regions of WM, GM, and CSF from the abnormalities visible in the vicinity of the primary tumor site---the peritumoral edematous/invaded tissue (\acr{ED}) and the tumor core region (\acr{TU}), which can be further differentiated into enhancing tumor (\acr{EN}) and the union of the necrotic and non-enhancing parts (\acr{NE}). An accurate segmentation of tumor sub-regions is relevant for diagnosis and treatment planning. However, tumor segmentation is quite challenging due to the tumor regions being defined through intensity changes relative to the surrounding normal tissue, as well as varying intensity distributions disseminated across multiple modalities, that are often subtle. Additional factors are imaging artifacts such as noise, motion, or magnetic field inhomogeneities. The manual annotation of region boundaries is time consuming, and prone to misinterpretation, human error and observer bias~\cite{Deeley:2011a,Mazzara:2004a}. To remove these biases it is desirable to design automatic approaches. To date, the best results are achieved by machine learning techniques~\cite{Bakas:2018a, Isin:2016a}. These methods can be augmented with biophysical simulations~\cite{Angelini:2007a, Bauer:2013a}. For example, one can use simulations as a data augmentation strategy for training neural networks~\cite{Gholami:2019a}.

In \emph{image registration}, we are interested in computing a spatial transformation $\vect{y}$ that maps points in one image to its corresponding points in another image. Image registration has evolved to an indispensable tool in medical image analysis~\cite{Sotiras:2013a}. In the context of monitoring disease progression or treatment response, images of a brain tumor patients will be acquired at different points in time with changes in morphological appearance, texture, structure, shape of the tumor, and the field of view~\cite{Bauer:2013a, Mang:2008a}. These changes make an accurate registration a delicate matter. While changes in pose can be compensated for in a stable way~\cite{Mang:2008a}, it is challenging to compensate for non-linear deformations $\vect{y}$~\cite{Mang:2007c}. This is in particular the case when it comes to registration between pre- and post-resection imaging scans~\cite{Kwon:2014a,Han:2019:prePostRegis}, where parts of the tumor are missing due to resection. Aside from monitoring a single patient, we might be interested in gathering statistical information across a population of patients~\cite{Ellingson:2013:atlasPhenotypes, Bilello:2016:atlas}. This necessitates the registration of patient individual scans to a common anatomical atlas image, which---even in the absence of a tumor---is challenging due to inter-patient anatomical variability. In the presence of a tumor, this registration requires finding correspondences between two topologically different images---one with and one without tumor. Similarly, we can use image registration to solve the segmentation problem by transferring labels for anatomical regions defined in the atlas space to unseen patient data~\cite{Gooya:2012a, Scheufele:2018a, Zacharaki:2009a, Zacharaki:2008a}. A simple strategy is to mask the tumor area (known as cost-function masking)~\cite{Stefanescu:2004a, Henn:2004a, Brett:2001a}, or to relax the registration in the area affected by the tumor~\cite{Parisot:2014a}. This yields poor registration quality for tumors with severe mass effect. Another strategy is to simultaneously invert for the deformation map and a drift in intensity representing the imaging abnormality~\cite{Li:2012a}. While this may produce acceptable results for the purpose of atlas-based segmentation and registration, it cannot be used for model prediction and tumor characterization~\cite{Ellingson:2013:atlasPhenotypes, Akbari:2018:egfrviii}---our ultimate goal. One remedy is the integration of image registration with biophysical modeling.

\subsection{Integration of Biophysical Modeling with Imaging}

In aggressive tumors, time series of images of patients that have not undergone treatment are in general not available (this is different in animal studies~\cite{Hormuth:2015a, Lima:2017a, Feng:2018a}). The fact that we do not have access to longitudinal data without treatment makes the integration of biophysical modeling with imaging even more challenging. We need to calibrate complex PDE models (see \secref{s:tumor-modeling}) based on a single snapshot in time. Since the model is typically a dynamical system it is impossible (without regularization) to calibrate model parameters using a single-time dataset. Moreover, we do not have data for the initial state, i.e., an image of the patient's brain without a tumor, or any other information about time history. A common strategy is to simulate the progression of the tumor in a healthy (tumor-free) image (see \figref{f:glistr:results}).

This is sub-optimal since anatomical differences introduce significant errors~\cite{Mang:2014a,Mang:2012b}. One approach for resolving these anatomical differences is to simultaneously invert for a model-based deformation (spatial transformation) $\vect{y}$ that maps the atlas to the patient anatomy (or vice versa)~\cite{Gooya:2012a, Scheufele:2018a, Zacharaki:2009a, Hogea:2008a, Mang:2012c, Scheufele:2019a, Mohamed:2006a, Zacharaki:2008a, Zacharaki:2008b}. The models described in~\cite{Zacharaki:2009a, Mohamed:2006a, Zacharaki:2008a, Zacharaki:2008b} are oversimplified (e.g., purely mechanical); they do not allow for recovering growth patterns of tumors with complex shapes, nor do they provide information about progression or infiltration of tumor cells into healthy tissue. We discuss two frameworks developed by our group that do not have these limitations. They integrate complex biophysical simulations with image registration in an attempt to aid imaging workflows and provide predictive capabilities. The first framework is the \emph{GLioma Image SegmenTation and Registration} (\acr{GLISTR})~\cite{Gooya:2012a,Bakas:2016a}. The second one is the \emph{Scalable Integrated Biophysics-based Image Analysis} (\acr{SIBIA})~\cite{Scheufele:2018a, Gholami:2017a, Scheufele:2019a, Mang:2017c}. Variants of these frameworks are already used in clinically-relevant studies~\cite{Gooya:2012a, Binder:2018:CancerCell, AdvancingTCIA, Yankeelov:2013a, Menze:2011a, Gooya:2011a} (see \secref{s:clinically-relevant-studies}).

\subsubsection{GLioma Image SegmenTation and Registration (GLISTR \& GLISTRboost)}

GLISTR~\cite{Gooya:2012a} is a generative approach for simultaneously registering a probabilistic atlas of a healthy population to brain tumor \mpmri{} scans, and segmenting the apparent brain in various sub-regions. The output of GLISTR is a posterior probability map $\pi_i : \Omega \to [0,1]$, $i \in \Theta$ and a label map $\xi : \Omega \to \Theta$, $\Theta \defeq \{\text{W},\text{G},\text{TU},\text{F},\text{ED}\}$. GLISTR incorporates the glioma growth model described in~\cite{Hogea:2008a,Hogea:2007a,Hogea:2008b} (see also \secref{s:tumor-modeling}). We define the probability maps $\pi_i$ as conditional probabilities $\pi_i (\vect{x}\!\mid\!\vect{p})$ on the \emph{unknown} tumor parameters $\vect{p}$.

\begin{figure}
\begin{center}
\includegraphics[width=0.98\textwidth]{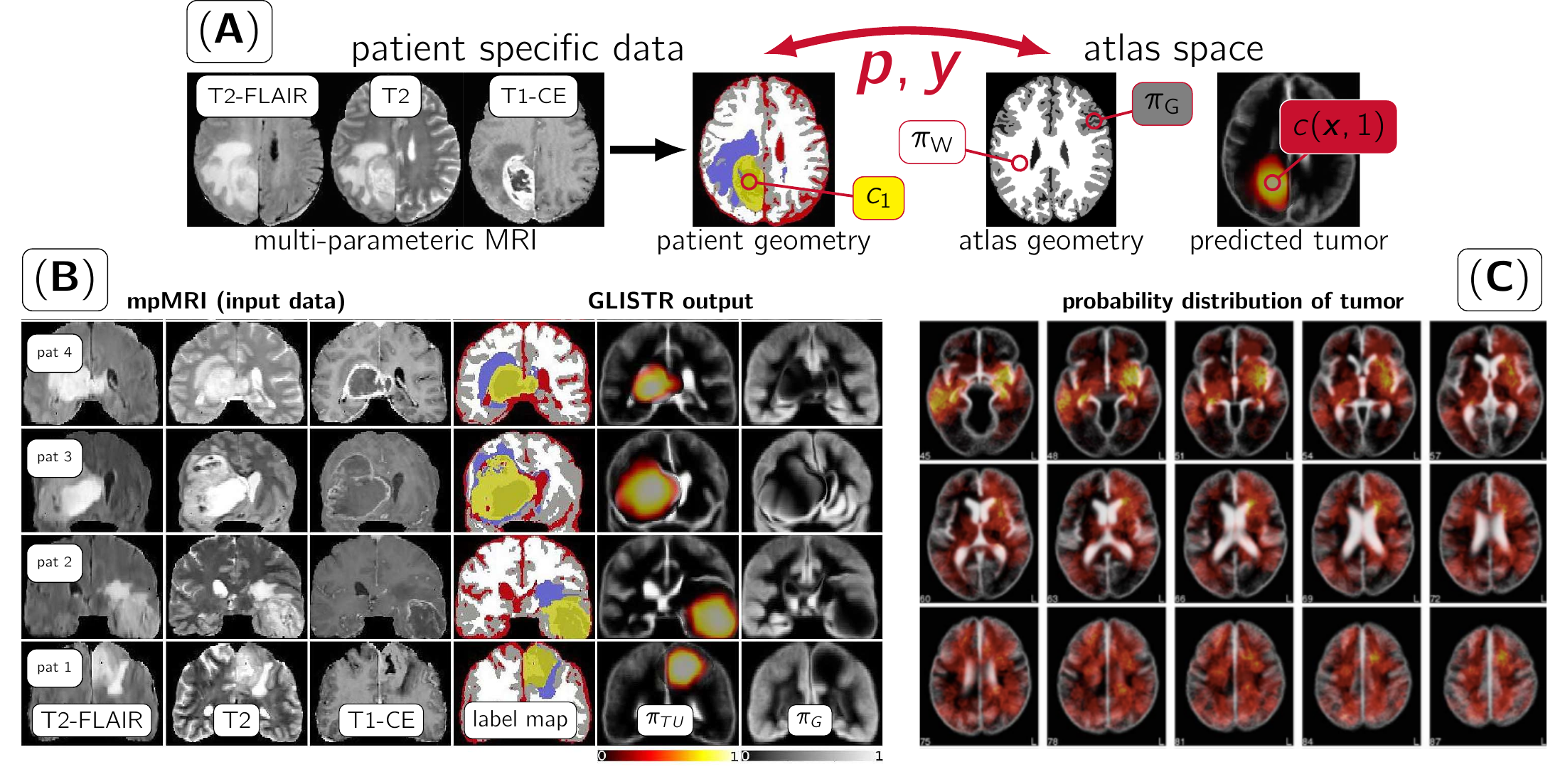}
\caption{({\bf A}) Illustration of the inverse problem of estimating patient specific model parameters $\vect{p}$. We seek parameters $\vect{p}$ such that the predicted state $c(\vect{x},1)$ (solution of the forward problem) matches some observed data $c_{\text{OBS}}$. The input data to our problem are \mpmri{} data (top left). The image in the middle illustrates data we present to our solver. The image on the right shows the model output for the computed parameters. The simulations are performed in a tumor-free atlas image (second column to the right). To compensate for anatomical differences in patient and atlas geometry, we additionally invert for a deformation map $\vect{y}$. ({\bf B}) and ({\bf C}) Exemplary results for GLISTR~\cite{Gooya:2012a}. We show segmentation results (in(see ({\bf B}); coronal planes) and tumor probability maps (see ({\bf C}); axial planes). ({\bf B}) Each row corresponds to a different patient (bottom to top: pat 1 through pat 4). The input \mpmri{} data is shown on the left. The three images to the right show the computed tumor labels $\xi$ (EN: light yellow; NE: dark yellow; ED: purple; CSF: red; GM: gray; WM: white), the probability map for the tumor $\pi_{\text{TU}}$, and the probability map of GM $\pi_{\text{G}}$. ({\bf C}) shows the average of the computed tumor posteriors across 122 glioma cases. The color map is the same as the one used for $\pi_{\text{TU}}$. It can be seen that within the considered patient population, the region with the highest tumor probability is placed in the left temporal lobe of the brain. (Figure adapted from References~\cite{Mang:2018a,Gooya:2012a}. ({\bf A}) reprinted by permission from Springer Nature, Optimization and Engineering, Copyright \textcopyright2018 Springer. ({\bf B}) and ({\bf C}) reprinted by permission from IEEE, IEEE Transactions on Medical Imaging, Copyright \textcopyright2012 IEEE).\label{f:glistr:results}}
\end{center}
\end{figure}

The joint registration and segmentation problem solved in GLISTR is as follows: We are given a vector $\vect{q}(\vect{x}) \defeq (q_1(\vect{x}),\ldots,q_k(\vect{x}))\in\ns{R}^k$ of observations (imaging intensities) that correspond to $k$ MRI modalities (input to our problem; see \figref{f:glistr:results}). We seek a deformation map $\vect{y}$, model parameters $\vect{p}$, and the intensity distributions $\vect{\phi}$ (mean and covariance of a Gaussian distribution; see below) for the labels $\Theta$. The deformation map $\vect{y}$ defines the registration between the patient specific image and the atlas space. The model parameters inverted for are given by $\vect{p} \defeq \left\{\vect{x}_0,\gamma,\kappa_{W},T\right\}$ with predefined initial condition $\pi_{TU}(\vect{x},0) \propto \exp(\|\vect{x}-\vect{x}_0\|_2^2)$, initial seed location $\vect{x}_0 \in \Omega$, diffusion coefficient $\kappa_{W} > 0$ for WM, final time $T>0$, and $\gamma >0$ determines the strength of the tumor mass effect (see \figref{f:forward-models} for an illustration). Under the assumption that the conditional probability distribution function of each $\vect{q}(\vect{x})$ can be modeled as a weighted mixture of Gaussians, we can solve for ${\vect{\phi},\vect{y},\vect{p}}$ as follows:
\begin{equation}
\label{e:glistr:opt}
\maxopt_{\vect{\phi},\vect{y},\vect{p}}\;\;
\textstyle\prod_{\vect{x}\in\Omega}\sum_{i \in \Theta}
\log(\pi_i (\vect{y}(\vect{x})\!\mid\!\vect{p}) g_i(\vect{q}(\vect{x})\!\mid\!\vect{\phi})),
\end{equation}

\noindent where $g_i(\vect{q}(\vect{x})\!\mid\!\vect{\phi})$ is a multivariate Gaussian distribution with mean $\vect{\mu}_i$ and covariance matrix $\mat{\Sigma}_i$, and $\vect{\phi} \defeq \{\vect{\mu}_i,\mat{\Sigma}_i\}_{i\in \Theta}$. To compute the maximum a posteriori (\acr{MAP}) estimate of~\eqref{e:glistr:opt} an expectation-maximization approach is considered. The optimization problem is solved using a derivative-free algorithm. We showcase results obtained with GLISTR in \figref{f:glistr:results}.

GLISTRboost~\cite{Bakas:2016a}, an extension of GLISTR, is a hybrid generative-discriminative model. The generative model is GLISTR; the discriminative part is based on a voxel-level multi-class classification through a gradient boosting ensemble model of decision trees~\cite{Friedman:2001a, Friedman:2002a}. It refines the tumor labels obtained from solving~\eqref{e:glistr:opt} based on information from multiple patients. The classifier is trained using the glioma data of the brain tumor segmentation (\acr{BraTS}) challenge~\cite{Bakas:2018a, AdvancingTCIA, Menze:2015a, tcgaGbmSegmentations, tcgaLggSegmentations}. Decision trees of maximum depth three are trained in a subset of the data to introduce randomness. A cross-validation framework is used to avoid overfitting. Additional randomness is introduced by sampling stochastically a subset of imaging features at each node. The features used for training the discriminative part of GLISTRboost consist of five components: \bipa\item image intensities from individual \mpmri{} and their differences across all of the datasets, \item first and second order image derivative information, \item the geodesic distance transform~\cite{Gaonkar:2015:geodesic} from an initialized seed-point $\vect{x}_0$ of the tumor, \item the posterior probability maps $\pi_i$, and \item first and second order texture statistics computed from a gray-level co-occurrence matrix\eipa. In a last step, each segmentation is refined by assessing the local intensity distribution of the current segmentation labels across each patient's \mpmri{} scans and updating their spatial configuration based on a Bayesian probabilistic framework~\cite{Bakas:2017a}. Here, the intensity distributions of the WM, ED, NET and ET are populated, considering the corresponding voxels of tissue probability equal to one. The histograms of the three pair-wise distributions considered (i.e., ED vs WM in T2-FLAIR, ET vs ED in T1-CE, and ET vs NET in T1-CE) are then normalized. The maximum likelihood estimation is then used to model the class-conditional probability densities ($\operatorname{Pr}(I(v_{i}) | \mathrm{class}$)) of each class by a distinct Gaussian model for each class. The voxels of each class in a close proximity (offset = 4) to the voxels of the paired class, were then iteratively evaluated by assessing their intensity $I(v_{i})$ and comparing the $\operatorname{Pr}(I(v_{i})|\mathrm{class}_{1}$) with $\operatorname{Pr}(I(v_{i})|\mathrm{class}_{2}$). The voxel $v_{i}$ was then classified into the class with the larger conditional probability, which is equivalent to a classification based on Bayes' theorem with equal priors for the two classes, i.e., $\operatorname{Pr}(\mathrm{class}_{1}) = \operatorname{Pr}(\mathrm{class}_{2}) = 0.5$. GLISTRboost has been evaluated on the testing datasets ($n=186$) of the BraTS 2015 challenge and was ranked as the top-performing method~\cite{Menze:2015a,Bakas:2016a}.

\subsubsection{Scalable Integrated Biophysics-based Image Analysis (SIBIA)}

SIBIA~\cite{Scheufele:2018a, Gholami:2017a, Scheufele:2019a, Mang:2017c} is a novel framework for integrating bio-physical simulations with \mpmri{} and optimization. It is a continuation of our efforts described in~\cite{Gooya:2012a, Zacharaki:2009a, Bakas:2016a, Hogea:2008a, Zacharaki:2008a}. It addresses several limitations of GLISTR/GLISTRboost, the main one being the need to manually select the tumor seed $\vect{x}_0$. SIBIA is fully automatic and does not require user intervention. Like GLISTR/GLISTRboost, SIBIA uses \bipa\item the biophysical models described in~\cite{Gholami:2016a,Subramanian:2019a} for the tumor modeling part (see also \secref{s:tumor-modeling}) and \item the \emph{Constrained LArge deformation diffeomorphic Image REgistration} package (\acr{CLAIRE})~\cite{Mang:2015a,Mang:2016b,Mang:2019a} as a module for diffeomorphic registration\eipa. The formulation is in spirit similar to~\cite{Hogea:2008a}. We invert for a velocity field $\vect{v}$ that parameterizes the deformation map $\vect{y}$ from the patient to the atlas space and tumor model parameters $\vect{p}$ using a PDE-constrained formulation of the form~\cite{Scheufele:2018a}
\begin{subequations}
\label{e:sibia:opt}
\begin{align}
&\minopt_{\vect{p},\vect{v},\vect{m},c}\;
  \half{1}\int_{\Omega}\! (c_{\text{A}}(\vect{x},1)\!-\!c_{\text{P}}(\vect{x},1))^2\!\d \vect{x}
+ \half{1}\int_{\Omega}\! \|\vect{m}_{\text{A}}(\vect{x},1)\!-\!\vect{m}_{\text{P}}(\vect{x},1)\|_2^2\d \vect{x}
+ \fs{R}(\vect{\vect{p},\vect{v}})\\
&\text{subject to }\; \D{F}_T(c,\vect{m},\vect{v},\vect{p}) = 0, \quad \D{F}_R(c,\vect{m},\vect{v}) = 0.
\end{align}
\end{subequations}

\noindent This model is a direct extension of the formulations in \secref{s:inversion}. The functions $c_{\text{A}}(\vect{x},1)$, $c_{\text{P}}(\vect{x},1)$, $\vect{m}_{\text{A}}(\vect{x},1)$ and $\vect{m}_{\text{P}}(\vect{x},1)$ are tumor and tissue probability maps defined in the patient (P) and atlas (A) space. The input data to the inverse problem are estimates for \bipa\item patient-specific tumor probabilities $c_{\text{OBS}} : \Omega \to [0,1]$, \item patient-specific material properties $\vect{m}_{\text{OBS}}: \Omega\to[0,1]^3$, $\vect{m}_{\text{OBS}}(\vect{x}) \defeq (\pi_{\text{OBS},\text{W}}(\vect{x}),\pi_{\text{OBS},\text{G}}(\vect{x}),\pi_{\text{OBS},\text{F}}(\vect{x}))$ (\emph{patient geometry}), and \item for the tumor-free patient-geometry (the atlas image) $\vect{m}_{\text{ATL}}: \Omega\to[0,1]^3$, $\vect{m}_{\text{ATL}}(\vect{x}) \defeq (\pi_{\text{ATL},\text{W}}(\vect{x}),\pi_{\text{ATL},\text{G}}(\vect{x}),\pi_{\text{ATL},\text{F}}(\vect{x}))$ (\emph{atlas geometry})\eipa. The operator $\D{F}_T$ is the forward tumor model (see~\eqref{e:fwd:multispecies} for an example); $\D{F}_T$ is used to predict a tumor in the atlas space that best matches the patient's tumor. The operator $\D{F}_R$ is the forward registration model (a hyperbolic transport equation); it is used to map the patient data to the atlas space. The input to the forward registration operator $\D{F}_R$ is the patient-specific tumor and tissue probabilities $c_{\text{OBS}}(\vect{x})$ and $\vect{m}_{\text{OBS}}(\vect{x})$, respectively.

Formally, our scheme proceeds as follows: Given some trial tumor parameter $\vect{p}$, the forward tumor model $\D{F}_T$ produces the predicted tumor probabilities $c_{\text{A}}(\vect{x},1)$ and tissue probability maps $\vect{m}_{\text{A}}(\vect{x},1) = (\pi_{1,\text{W}}(\vect{x}),\pi_{1,\text{G}}(\vect{x}),\pi_{1,\text{F}}(\vect{x}))$ at time $t=1$ in \emph{the atlas space}, where $\pi_{1,j}(\vect{x}) \defeq \pi_{\text{ATL},j}(\vect{x})(1-c_{\text{A}}(\vect{x},1))$, $j\in\{\text{W},\text{G},\text{F}\}$ are the updated probability maps for WM, GM, and CSF (healthy brain anatomy). Given some trial velocity field $\vect{v}(\vect{x})$, the forward registration model $\D{F}_R$ generates a spatially-transformed representation $c_{\text{P}}(\vect{x},1)$ and $\vect{m}_{\text{P}}(\vect{x},1)$ at pseudo-time $t=1$ of the patient specific data $c_{\text{OBS}}(\vect{x})$ and $\vect{m}_{\text{OBS}}(\vect{x})$. In the inverse problem~\eqref{e:sibia:opt}, we seek control variables $\vect{p}$ and $\vect{v}$ such that the tumor and tissue probability maps $c_{\text{A}}(\vect{x},1)$, $c_{\text{P}}(\vect{x},1)$, $\vect{m}_{\text{A}}(\vect{x},1)$ and $\vect{m}_{\text{P}}(\vect{x},1)$ defined in the atlas and patient space are as close as possible. We measure the proximity between these data using a squared $L^2$-distance in \eqref{e:sibia:opt}. The functional $\fs{R}$ in \eqref{e:sibia:opt} is a regularization model for the control variables $\vect{p}$ and $\vect{v}$.

Computing the minimizer of \eqref{e:sibia:opt} is conceptually equivalent to computing the MAP point for \eqref{e:glistr:opt}. In SIBIA, we invert for the growth rate $\rho > 0$, the diffusivity $\kappa > 0$ and/or the initial condition $c_0(\vect{x})$, where $c_0(\vect{x}) \defeq \sum_{k=1}^r w_k\phi_k(\vect{x})$ is modeled as an $r$-dimensional space spanned by Gaussian basis functions $\phi_k : \Omega \to \ns{R}$. This parameterization allows us to model multi-focal and multi-centric tumors. SIBIA~\cite{Scheufele:2018a, Gholami:2017a, Scheufele:2019a, Mang:2017c} uses a globalized, adjoint-based method (i.e., derivatives of the Lagrangian). We do not iterate on both control variables $\vect{v}$ and $\vect{p}$ simultaneously. We perform a block elimination instead, and iterate, resulting in an interleaved optimization on the controls exploiting dedicated solvers for the individual sub-blocks~\cite{Gholami:2016a,Gholami:2017a,Mang:2015a,Mang:2019a}. SIBIA has been deployed in parallel computing platforms to further amortize computational costs~\cite{Gholami:2017a, Mang:2016b, Mang:2019a}.

We have applied SIBIA to hundreds of real 3D datasets and achieved encouraging results for atlas-based segmentation~\cite{Scheufele:2018a,Mang:2017c}. However, our initial scheme~\cite{Scheufele:2018a} does not allow for reliably inverting for meaningful model parameter $\vect{p}$; its predictive capabilities are limited. One key issue is that we map the patient geometry to the atlas space. In our most recent work~\cite{Scheufele:2019a} we have changed the formulation to map the atlas geometry to the patient space, excluding tumor probabilities. We hypothesize (and demonstrated experimentally through synthetic test problems~\cite{Scheufele:2019a}) that this improved scheme---in combination with a sparsity soft constraint for the parameterization of the initial condition~\cite{Subramanian19aL1}---allows us to more reliably invert for patient-specific tumor parameters.

\section{CLINICALLY RELEVANT STUDIES}
\label{s:clinically-relevant-studies}

Numerous clinically-relevant studies revolve around precision diagnostics~\cite{Davatzikos:2019:precisionDiagnostics} leveraging rich information from biophysical models of tumor growth~\cite{Tracqui:2009}. Considering the complexity of routinely acquired advanced \mpmri\ of GBM patients~\cite{Shukla:2017:advancedMri}, there is an apparent need for advanced computational algorithms for automated image analyses. Such analyses include automated brain tumor segmentation algorithms coupled with biophysical growth models~\cite{Gooya:2012a, Scheufele:2018a, Bakas:2016a, Scheufele:2019a, Kwon:2014a, Gooya:2011a, Zeng:2016:segmentation, Kwon:2014:MICCAI, Kwon:2014b} (see also \secref{s:integration-with-mri}) leading to an accurate quantitative assessment of the distinct histologically heterogeneous tumor sub-regions, potentially benefiting the clinical workflow in radiology and radiation oncology settings, as well as providing platforms for radio(geno)mic research. There are ample literature examples that support the benefit of biophysical tumor growth modeling contributing towards robust patient-specific tumor characterization (i.e., \emph{personalized medicine}), especially by virtue of accurate population-based spatial distribution atlases of GBM~\cite{Bilello:2016:atlas}.

We appreciate the potential benefit of incorporating biophysical modeling in clinical research studies, and to facilitate the widespread use of biophysical modeling, we have already integrated GLISTR and GLISTRboost into the extended suite of the \emph{Cancer Imaging Phenomics Toolkit} (\acr{CaPTk})~\cite{Davatzikos:2018:captk, Rathore:2017:brainCaptk}, and made them available for public use through the online \emph{Image Processing Portal} (\acr{IPP}; further details can be found at \href{https://ipp.cbica.upenn.edu}{https://ipp.cbica.upenn.edu}) of the \emph{Center for Biomedical Image Computing and Analytics} (\acr{CBICA}; see \href{https://www.cbica.upenn.edu}{https://www.cbica.upenn.edu}). CBICA's IPP allows users to perform their analyses without any software installation through CBICA's computing resources. The integration of CLAIRE and SIBIA with CaPTk is an ongoing project. The sections below summarize a few example studies relating radiographic analyses to specific endpoints (e.g., clinical outcome, molecular characteristics) and do not intend to be a complete literature review.

\subsection{Prediction of Patient Overall Survival}

\begin{figure}
\centering
\includegraphics[width=\textwidth]{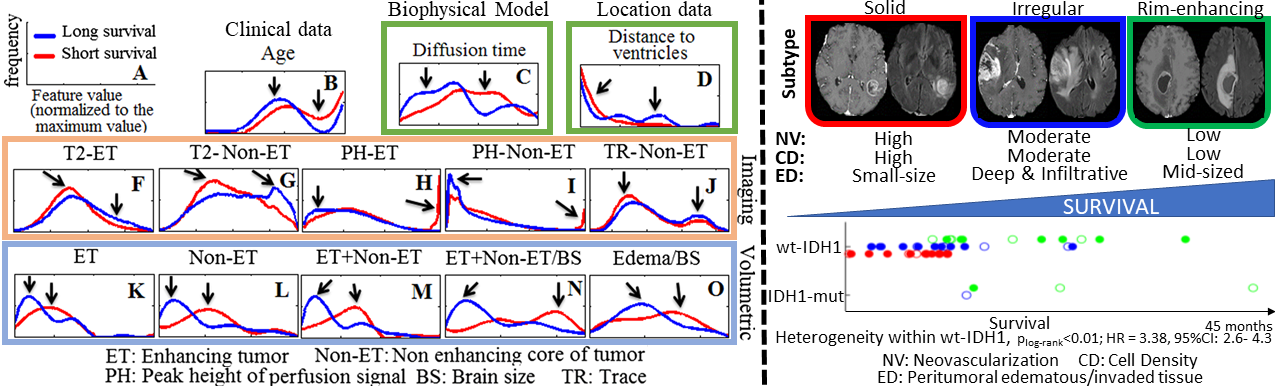}
\caption{Example studies on predicting patient OS. Left: distributions of features most predictive of OS across long- (blue) and short- (red) survivor groups. The black arrows point to larger differences between the groups, per feature. Notably, the diffusion time obtained via biophysical models of tumor growth is one of the most distinctive features. Right: distinction of radiographic subtypes in relation to patient OS. The shortest survival of the IDH-mut, belonged to the irregular subtype, which overall had lower OS, indicating that the radiographic subtype can potentially add predictive value within IDH1-mut patients. (Figures modified from~\cite{macyszyn.survival,Rathore:2018:SciRep}. Reprinted by permission from Oxford University Press, Neuro-Oncology, Copyright \textcopyright2016 Oxford University Press and from Springer Nature, Scientific Reports, Copyright \textcopyright2018 Springer.)}
\label{f:clinical_survival}
\end{figure}

Patient overall survival (\acr{OS}) is the ultimate clinical outcome; accurate predictions could affect clinical decision-making and treatment planning. Numerous studies have been focusing on GBM prognostic evaluation and stratification. These studies support the benefit of incorporating biophysical growth models~\cite{macyszyn.survival, rathore.survival, Rathore:2018:SciRep, Baldock:2014:survivalBenefit, Swanson:2008a, Wang:2009:survivalKinetics}, show the generalization across multi-institutional data~\cite{Rathore:2018a:interScanner, Rathore:2018b:interScanner}, even when being compared with the prognostic value of current clinical and genomic markers, and demonstrate that an integration of models with imaging offers additive prognostic value even beyond the current WHO classification~\cite{Rathore:2018:beyondIdh, Rathore:2018:SciRep, Shukla:2017:Astro} (see \figref{f:clinical_survival}). Furthermore, integration of biophysical growth modeling with advanced radio-phenotypical features derived from basic structural \mpmri\ (i.e., T1, T1-CE, T2, T2-FLAIR) can compensate for the lack of advanced \mpmri\ scans (e.g., DSC-MRI, DTI), and still offer comparable prognostic predictions~\cite{Bakas:2017:snoAdvancedFeatsAdvancedModalities}.

\subsection{Treatment Planning}

Although more than 90\% of tumor recurrence occurs within ED~\cite{Petrecca:2013}, there is limited research focused on its assessment and its microenvironment~\cite{Lemee:2015:peritumoral}. ED appears to develop in response to angiogenic and vascular permeability factors associated with infiltrating tumor~\cite{Chang:2007}. As tumors outgrow the native blood supply, the resultant ischemia triggers further secretion of angiogenic factors that promote vascular proliferation~\cite{Bullitt:2005,Kerbel:2000}. Advanced computational analyses, incorporating biophysical tumor growth modeling, have been conducted to evaluate the amount of heterogeneous tumor infiltration in the ED tissue and by that assess the risk of recurrence~\cite{akbari.recurrence, akbari.recurrence2, Rathore:2018:recurrence}. The value of these studies has been retrospectively validated in independent discovery and replication cohorts and has shown significant results (odds ratio $>13$). Furthermore, they were recognized for their usefulness as potential therapeutic tools and are currently under a clinical trial setting of targeted personalized dose escalation planning.

Along the same lines, various other studies have attempted to shed light on the quantification of microscopic tumor invasion and cell proliferation~\cite{Ellingson:2011:invasionQuantification, Stein:2007:invasionMathematicalModel}, and also tumor growth rates in relation to diffusion~\cite{Burgess:1997:growthRates}, while incorporating growth modeling.

\subsection{Radiogenomics: Non-invasive Tumor Molecular Characterization}

Current tumor molecular characterization is based on \emph{ex vivo} tissue analysis that cannot capture the tumor's spatial heterogeneity. Since radiographic imaging is routinely acquired and can capture the whole tumor extent, multiple studies have focused on non-invasive prediction of molecular characteristics of GBM from radiographic tumor patterns~\cite{Rathore:2019:aacr} while incorporating biophysical growth modeling; from genome wide association analysis with tumor spatial distribution patterns~\cite{Elsheikh:2018:gwasGbm}, to individual molecular alterations~\cite{Rathore:2018:mgmtSno, Rathore:2018:mgmtJco, bakas.egfrviii, Bakas:2018:idh, Akbari:2018:egfrviii}, as well as transcriptomic GBM subtypes \cite{macyszyn.survival, Verhaak:2010:subtypes}. Below we provide a few example studies focusing on non-invasive prediction of ones of the most important GBM molecular characteristics; the isocitrate dehydrogenase-1 (IDH1); the O6-methylguanine DNA methyltransferase (MGMT); and the Epidermal Growth Factor Receptor variant III (EGFRvIII).

According to the 2016 WHO classification \cite{Louis:2016a}, determination of the IDH1 mutational status is essential for the clinical diagnosis and treatment planning of glioma. The ability to identify IDH1 at initial patient presentation can influence decision-making and appropriate treatment planning. Furthermore, as IDH1 mutant (IDH1-mut) enzyme inhibitors and immunotherapeutic options are developed, non-invasive determination at pre-operative and follow-up timepoints can be influential. With that in mind, a pre-operative non-invasive signature of IDH1 was constructed based on quantitative radiographic phenotypical features from a retrospective cohort of 86 high grade glioma \mpmri\ scans (IDH1-mut:15)~\cite{Bakas:2018:idh}. The features integrated volumetric and morphological measurements, texture descriptors, location characteristics, and biophysical growth model parameters~\cite{Bakas:2016a}. Following multivariate cross-validated forward-sequential feature selection, 61 of these features were identified as the most discriminative, primarily including texture descriptors and a distinct spatial location of the IDH1-mut tumors with more prominence in the frontal/occipital lobe. Quantitative evaluation of this signature yielded an accuracy of 88.4\% (sensitivity=66.7\%, specificity=92.9\%, AUC=0.81) on classifying IDH1 mutational status.

MGMT promoter methylation is another well-accepted prognostic indicator in GBM that directly influences the effectiveness of chemotherapy, where specifically methylated tumors (MGMT\texttt{+}) are more responsive. A non-invasive signature for the status of the MGMT promoter methylation could contribute in addressing limitations of the current determination, which can be limited by inadequate tissue specimen or assay failures. In~\cite{Rathore:2018:mgmtSno,Rathore:2018:mgmtJco} a retrospective cohort of 122 patients diagnosed with pathology-proven \emph{de novo} GBM (MGMT\texttt{+}:46) and available pre-operative \mpmri\ scans was identified. 330 radiographic phenotypical features were extracted per patient, including measurements of volume, morphology, texture, and voxel-wise location characteristics obtained after incorporating a biophysical model~\cite{Bakas:2016a}. Multivariate cross-validated forward sequential feature selection was applied to identify 46 features as the ones to create the non-invasive signature, which revealed that MGMT\texttt{+} tumors have lower neovascularization and cell density compared with MGMT-unmethylated tumors. Assessment of the location characteristics yielded a distinctive spatial pattern, with the MGMT\texttt{+} tumors being lateralized more to the left hemisphere compared with MGMT-unmethylated tumors. The cross-validated accuracy of this signature in correctly classifying MGMT\texttt{+} tumors was 84.43\% (sensitivity=80.43\%, specificity=86.84\%, AUC=0.85).

\begin{figure*}
\centering
\includegraphics[width=\textwidth]{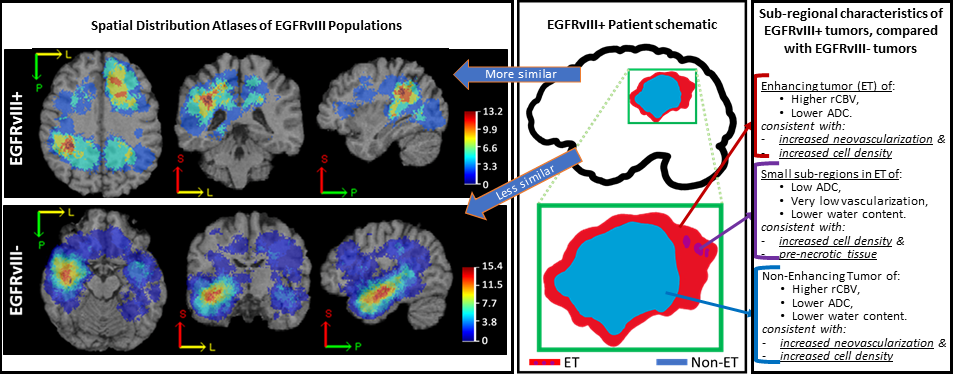}
\caption{Spatial descriptive characteristics of EGFRvIII glioblastoma, following advanced computational analysis incorporating biophysical tumor growth modeling. (Figure modified from \cite{Akbari:2018:egfrviii}. Reprinted by permission from Oxford University Press, Neuro-Oncology, Copyright \textcopyright2018 Oxford University Press.)}
\label{f:clinical_EGFRvIII}
\end{figure*}

Another GBM driver mutation is EGFRvIII, which has been considered in multiple GBM clinical trials. It was recently discovered that GBM harboring the mutation have a distinct spatial distribution pattern when compared to tumors without the mutation, that could distinguish a mutant tumor with $75\%$ accuracy~\cite{Akbari:2018:egfrviii}. \figref{f:clinical_EGFRvIII} shows the voxel-wise location prominence of GBM, in spatial distribution atlases relating to the presence/absence of the EGFRvIII mutation, after incorporating biophysical modeling~\cite{Bakas:2016a}. We note that there exists another pre-existing study evaluating the location prominence of GBM stratified by their EGFRvIII status, \emph{without} incorporating biophysical modeling parameters in their image analysis, and identifying different results~\cite{Ellingson:2013:atlasPhenotypes}. Notably, the studies that incorporated biophysical modeling~\cite{bakas.egfrviii, Akbari:2018:egfrviii}, evaluated a three times larger sample size, which in turn is expected to potentially provide more robust statistics of spatial patterns.

\subsection{Generating Hypothesis for Further Investigation}

\begin{figure*}
\centering
\includegraphics[width=\textwidth]{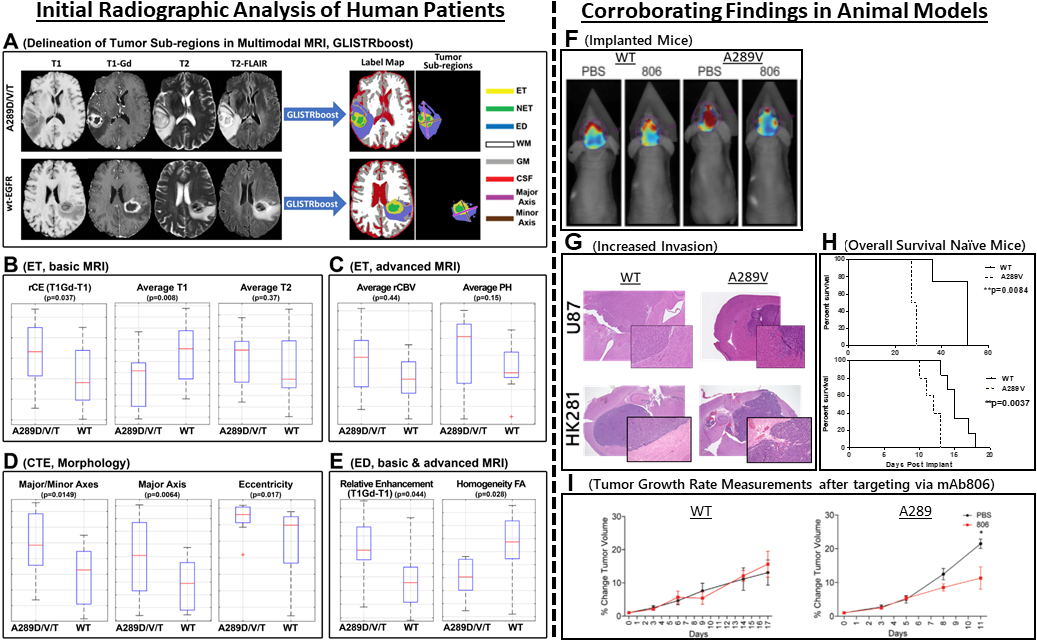}
\caption{Summary figure of computational radiographic analysis incorporating biophysical growth modeling (A-E) \cite{Bakas:2016a} leading to the discovery of a potential molecular target, presenting an opportunity for potential therapeutic development \cite{Binder:2018:SNO, Binder:2018:CancerCell}. The findings of the radiographic analysis were corroborated in mice implanted with tumors (F,H), the histological analysis of which (G) shows increased invasion. The implanted tumor growth rate was shown to be much decreased after targeting via mAb806. (Figure modified from \cite{Binder:2018:CancerCell}. Reprinted by permission from Elsevier, Cancer Cell, Copyright \textcopyright2018 Elsevier.)}
\label{f:clinical_A289}
\end{figure*}

Integrated analysis of advanced \mpmri\ scans and biophysical modeling~\cite{Bakas:2016a} has contributed on the discovery of a potential molecular target, presenting an opportunity for potential therapeutic development~\cite{Binder:2018:SNO, Binder:2018:CancerCell}. Radiographic signatures of EGFR extracellular domain missense mutants (i.e., A289V) were identified, suggestive of an invasive and proliferative phenotype~\cite{Binder:2018:CancerCell}, associated with shorter survival in patients. These findings were corroborated by experiments \emph{in vitro} and \emph{in vivo}---animal models. Kaplan-Meier survival curves comparing mice implanted with modified cell lines \emph{in vivo} (i.e., {U87} and {HK281} tumor cell lines expressing either wild-type EGFR or the EGFR A289V mutant---$n=6$ per group, $p<0.01$), demonstrated decreased overall survival, increased proliferation, and increased invasion. Further, mechanistic exploration revealed increased MMP1 expression (driven by ERK activation) leading to both increased proliferation and invasion. Finally, the tumor driver status of EGFR A289V was demonstrated by \emph{in vivo} targeting via an EGFR monoclonal antibody (mAb806), increasing animal survival and inhibiting tumor growth. These results serve to highlight the complexity of the EGFR signaling cascade and pathway nuances of extracellular domain mutations in the context of cancer \cite{Binder:2018:CancerCell}. \figref{f:clinical_A289} summarizes the results of this study \cite{Binder:2018:CancerCell}.

\section{CONCLUSIONS}
\label{s:conclusions}

We have reviewed existing approaches towards integration of computational models and image analysis for characterization of neuro-imaging data of brain tumor patients. We have described state-of-the-art technology for biophysical tumor growth modeling, and the inverse problem of estimating adequate parameters to fit the model output to available observations. We have discussed the integration of biophysical models with image analysis algorithms, and showcased clinically relevant results that demonstrate the benefit of such an integration.

Despite these encouraging results, we note that a successful integration of biophysical models with image analysis poses significant mathematical and computational challenges. First and foremost, biological systems involve complex, multi-faceted, heterogeneous biological, physical and chemical behavior at different spatial and temporal scales of observation~\cite{Rahman:2017a}. This makes the development of predictive models a difficult endeavor. Complex models result in a vast number of parameters~\cite{Rahman:2017a, HawkinsDaarud:2013a}, which makes them difficult to calibrate to medical imaging data, especially since clinical data provides only scarce information (e.g., single time point; imaging noise; low resolution; only indirect phenotypic measurements reflecting coarse aspects of these complex underlying biological processes). Aside from computational issues, additional mathematical and modeling issues such as \bipa\item uncertainties in the data and model; \item modeling errors and inadequate mathematical models; \item ill-posedness of the inverse problems (non-uniqueness of the solution); \item decisions about appropriate regularization models and data misfit terms; and/or \item influences of the numerical discretization on the inversion, make such an integration even more difficult\eipa. Consequently, other sources of information or, equivalently, strong modeling priors, need to be integrated to make these approach practical. One possibility to alleviate some of these challenges, and to define, test, analyze, and design appropriate priors, is through animal models or \emph{in vitro} studies, albeit this approach is well-recognized to be limited, as well, in its ability to generalize well to \emph{in vivo} cancer growth in humans.

Ample and diverse data are expected to contribute towards addressing these challenges and expedite further developments in biophysical modeling of the growth, invasion, and proliferation of untreated gliomas, as well as models of polyclonal gliomas following chemotherapy and surgical resection. Such data exist only across institutions, and the current paradigm for multi-institutional collaborations (i.e., pooling data in a centralized location) suffers from various privacy, technical, and ownership concerns. However, there are existing efforts on alternative collaboration paradigms based on distributed learning approaches~\cite{Sheller:2018, Chang:2018:distributed} that could be investigated further towards addressing the need for large datasets, while overcoming data-ownership concerns.

Integrated diagnostics increasingly demonstrate their clinical importance, with the most recent clinical example of the revised 2016 WHO classification for CNS tumors incorporating molecular characterization to histologic patterns~\cite{Louis:2016a}. However, several intrinsic and extrinsic factors hinder this molecular characterization, which currently requires \emph{ex vivo} invasive tissue analysis. Such analysis is limited in assessing the tumor's spatial heterogeneity and not amenable to relatively regularly repeated evaluations during treatment. On the other hand, \mpmri\ can non-invasively provide a macroscopic radiographic phenotype capturing the whole extent of a tumor. Since \mpmri\ scans of GBM patients are clinical routine (pre-operatively and longitudinally during adjuvant treatment), there is an opportunity for ample data to be utilized for developing dynamic non-invasive biomarkers. Our working hypothesis is that the integration of this data with sophisticated computational tools is beneficial for assessing the spatial and temporal heterogeneity of GBM, and has the potential to influence treatment, towards improving the health of GBM patients.

There is a notable growth of literature related to integrated radio-phenotypical diagnostics revolving around \emph{precision diagnostics}, i.e., the precise molecular characterization of tumors, by looking for patterns and targets identified from a population of patients. However, such non-invasive macroscopic integration, instead of revolving solely around \emph{precision medicine}, could also contribute on \emph{personalized}/\emph{adaptive} approaches that may expand on \emph{precision medicine} by characterizing within-patient heterogeneity, spatially and temporally. \emph{Personalized}/\emph{adaptive medicine} may have the potential to further customize treatment options using patient-specific factors. As tumor growth and invasion models become more elaborate, they might play a role in allowing to estimate patient-specific growth parameters that contribute to a more precise characterization of tumor properties.

Recent computational studies have provided evidence of non-invasive comprehensive multiscale characterization of the tumor's phenotype, behavior and microenvironment, before, during, and after treatment, thereby offering important information for diagnostic, prognostic, and predictive purposes, whilst capturing the whole extent and heterogeneity of a tumor. Integrated radio-phenotypical biomarkers may enable opportunities for non-invasive patient selection for targeted therapy, stratification into clinical trials, prognosis, and repeatable monitoring of molecular characteristics during the treatment course, leading to quantitative non-invasive evaluation of treatment response. Such advancements on integrated diagnostics, describing a composite multiscale index through synergistic analyses of radiographic, histopathologic, genetic, clinical, and biophysical data, may speed up scientific discovery and improve both \emph{precision} and \emph{personalized}/\emph{adaptive} medicine.

Based on current results, we are convinced that the integration of advanced computational, mathematical, and biophysical methods offers great promise to become an indispensable and influential tool for patient management. However, we acknowledge that there lies a significant amount of multi-disciplinary work ahead. Pending further clinical validation, we anticipate the integration of these tools into a future iteration of the WHO classification scheme for CNS tumors, thus providing a more complete understanding of the mechanisms of disease, leading to more effective treatment and beneficial patient prospects.

\section*{Acknowledgements} This work was partly supported by the Simons Foundation (586055); National Institutes of Health (5R01NS042645-14, R01NS042645, U24CA189523, UL1TR001878); ITMAT of UPenn; National Science Foundation (CCF-1817048, CCF-1725743, and DMS-1854853); U.S. Department of Energy, Office of Science, Office of Advanced Scientific Computing Research, Applied Mathematics program (DE-SC0019393); U.S. Air Force Office of Scientific Research (FA9550-17-1-0190). Computing time on the Texas Advanced Computing Centers’ (TACC) systems was provided by an allocation from TACC and the NSF.

\end{document}